\documentclass[9pt,twocolumn,twoside]{opticajnl}
\journal{opticajournal} 

\setboolean{shortarticle}{true}

\usepackage{lineno}

\title{Perturbed three-channel waveform synthesizer for efficient isolated attosecond pulse generation and characterization}

\author[1,2,3,\dag]{Dianhong Dong}
\author[1,2,3,\dag]{Hushan Wang}
\author[1,2,3,\dag\dag]{Bing Xue}
\author[1]{Kotaro Imasaka}
\author[1]{Natuski Kanda}
\author[2,3]{Yuxi Fu}
\author[1]{Yasuo Nabekawa}
\author[1,*]{Eiji J. Takahashi}

\affil[1]{RIKEN Center for Advanced Photonics, RIKEN, 2-1, Hirosawa, Wako-shi, Saitama, 351-0198, Japan}
\affil[2]{Center for Attosecond Science and Technology, Xi’an Institute of Optics and Precision Mechanics, Chinese Academy of Sciences, Xi’an, 710119, China}
\affil[3]{University of Chinese Academy of Sciences, Beijing 100049, China}
\affil[$\dag$]{These authors contributed equally to this work}
\affil[$\dag$$\dag$]{xuebing@opt.ac.cn}
\affil[*]{Corresponding author: ejtak@riken.jp}

\begin{abstract}
The generation of gigawatt-class isolated attosecond pulses (IAPs) is vital for attosecond pump-probe experiments. In such experiments, the temporal duration of IAPs must be determined quickly and accurately. In this study, we developed a perturbed three-channel waveform synthesizer for efficient IAPs generation and characterization at low repetition rates (~10 Hz). Intense IAPs centered at photon energies of 60 eV (227 as duration) in Ar and 107 eV (128 as duration) in Ne were generated by the driving field from a three-channel waveform synthesizer and characterized using all-optical frequency-resolved optical gating (AO-FROG), which accelerated the measurement time to several minutes, providing fast feedback for the tunability of the IAP source. The peak power of the IAPs is higher than that reported in the literature.
\end{abstract}

\setboolean{displaycopyright}{false} 

\begin{document}

\maketitle
Since isolated attosecond pulses (IAPs) were first demonstrated in 2001 \cite{hentschel_attosecond_2001}, the remarkable progress in the development of attosecond light sources and their applications have inspired breakthroughs in ultrafast science \cite{baltuska_attosecond_2003,chini_generation_2014,popmintchev_attosecond_2010}. IAPs are powerful tools for understanding ultrafast phenomena in atoms, molecules, and condensed matter. In particular, high-intensity IAPs enabled attosecond-pump–attosecond-probe experiments are considered as concise and reliable methods for studying ultrafast processes in matter \cite{fidler_nonlinear_2019,gonzalez-castrillo_quantum_2020,kang_author_2020,sansone_isolated_2006,li_attosecond_2020}.

In previous studies, we developed a custom-tailored carrier-to-envelope phase (CEP) stable multiterrawatt synthesized optical electric field with an operating frequency of 10 Hz to generate a sub-microjoule-class high-harmonic (HH) beam, which achieved extreme ultraviolet (XUV) supercontinuum corresponding to IAP \cite{takahashi_carrier-envelope_2015,xue_fully_2020}. A submicrojoule IAP with a central photon energy of 60 eV was demonstrated via frequency-resolved optical gating for the complete reconstruction of attosecond bursts (FROG-CRAB) \cite{mairesse_frequency-resolved_2005}, yielding a result of 226 as duration \cite{xue_gigawatt-class_2022}, demonstrating the realization of 1.1 gigawatt IAP on a tabletop. 

According to previous studies \cite{xue_gigawatt-class_2022,xue_custom-tailored_2021}, the output IAP duration can be tuned by adjusting the relative delay between the three channels to modify the synthesized waveform. However, to generate submicrojoule-class intense attosecond pulses, the repetition rate is often limited to low values, e.g., 10 Hz. Under such a low repetition rate, it takes more than 30 min to acquire data with a sufficient signal-to-noise ratio (SNR) for characterizing the IAP temporal profile using FROG-CRAB. Electron spectroscopy is a major reason for the long acquisition time because it is an indirect measurement of the spectrum via photoionization. The long acquisition time hampers the efficient feedback for precise tuning of output IAPs. Furthermore, although a supercontinuum reaching the soft X-ray region with energies of several tens of nanojoule has been reported \cite{xue_custom-tailored_2021}, its temporal duration has not been measured because of the lack of efficient optical components covering the continuum spectrum bandwidth (90–120 eV). 

In this study, we developed a perturbed three-channel waveform synthesizer by introducing an extra weak field, in which the IAPs are generated by waveform synthesizer, and are evaluated using AO-FROG \cite{yang_all-optical_2020}. Because this all-optical method is not limited by soft X-ray optics, we measured IAPs with photon energies of not only around 60 eV but also around 100 eV. Although the repetition rate was low (10 Hz), the evaluation time was only a few minutes due to the simplicity and efficiency; thus, the driver field condition in the three-channel optical synthesizer could be timely optimized. Furthermore, the shortest pulse duration of the generated IAPs was 128 as, indicating that the peak power of the generated soft X-ray IAPs exceeded subgigawatts. We also investigated the gating-field intensity dependence of the proposed method and found the practical limitation not for breaking the perturbation approximation.

The theoretical model of AO-FROG was proposed in previous studies \cite{yang_all-optical_2020,dudovich_measuring_2006}. To investigate the gating-field intensity dependence of the method, we analyzed the AO-FROG approach using relevant equations. According to the Lewenstein model \cite{lewenstein_theory_1994}, considering a driving pulse with a fundamental vector potential \(A_0(t)\), a weak perturbing pulse with a vector potential \(A_p(t,\tau)\), and relative delay $\tau$ between the two fields, the single-atom response dipole moment for a single trajectory under the two fields can be expressed as follows \cite{dudovich_measuring_2006}:
\begin{equation}
d(t,\tau)=d_0(t)e^{[-i(\sigma(t,\tau)]}+c.c.,
\label{eq1}
\end{equation}where $d_0(t)$ is the dipole moment without an extra perturbing field, and $\sigma(t,\tau)$ is the perturbing pulse-induced additional phase and is expressed as follows:

\begin{equation}
\sigma(t,\tau)=-\int_{t^{'}}^{t} [p-A_0(t^{''})]^{2}A_p(t^{''},\tau) \, dt^{''},
\label{eq2}
\end{equation}where $p$ is the canonical momentum. By performing the Fourier transform of the dipole acceleration, the delay-dependent evolution of the spectrogram can be expressed as follows:

\begin{equation}
\Omega(\omega,\tau)\propto\omega^4\ |\widetilde d_0(\omega)\otimes\mathcal{F}{\{e^{[-i\sigma(t,\tau)]}\}}|^2,
\label{eq3}
\end{equation} which has a similar format to that of conventional FROG in the frequency domain. For a homogeneous medium with a length L (small compared with the Rayleigh length of the driving laser) and low pressure (excluding the reabsorption), the intensity of the integrated spectra can be expressed as \(\Omega^\prime(\omega,\tau)\propto\Omega(\omega,\tau)sinc^{2}(\Delta k(\omega)L/2) \)
\cite{weissenbilder_how_2022}, where $\Delta k(\omega)$ is the phase mismatch factor, which depends on the degree of ionization under a fixed phase matching pressure. The change in the ionization probability due to the weak gating pulse can be ignored, and the phase mismatch term can thus be considered a constant. Therefore, the collective response of the medium has a similar form as the single-atom response. Hence, the temporal profiles of both $d_0(t)$ and perturbing field can be reconstructed from the measured trace using the retrieved algorithm. According to Equation (\ref{eq2}), the selection of the perturbation field is arbitrary; thus, the synthesized electric field can be possibly measured using AO-FROG. 

\begin{figure}[t]
\centering
\includegraphics[width=\linewidth, keepaspectratio]{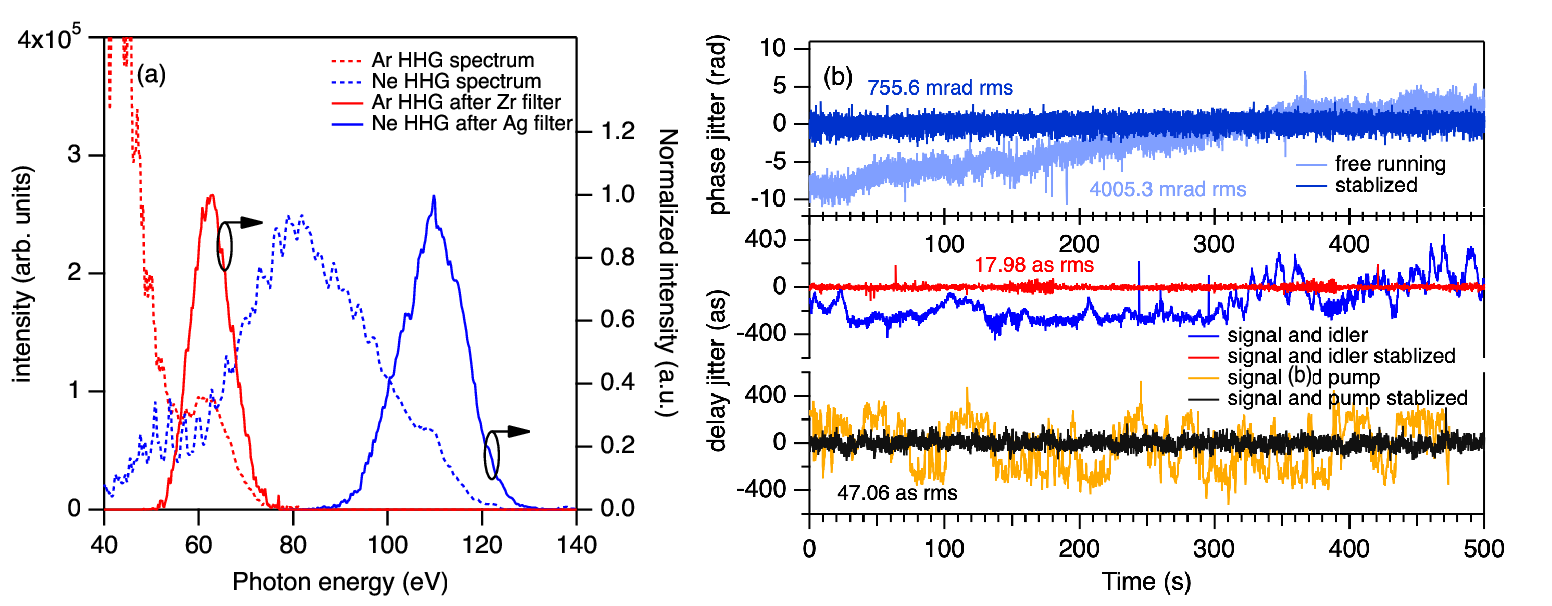}
\caption{(a) Calibrated accumulated supercontinuum intensity from Ar (red)/Ne (blue). Dash lines: without spectral filter; Solid lines: with metal filters. (b) Stability performance of the waveform synthesizer. CEP: 760 mrad rms; delay jitters: 47 as rms between signal and pump pulses; 18 as rms between signal and idler pulses.}\hrule\label{fig1}
\end{figure}

The XUV and soft X-ray HH beam, which generates IAPs, are driven by a custom-tailored CEP stable multiterrawatt optical electric field that includes three wavelength components from the OPA: the pump (800 nm, 30 fs, 21 mJ for argon, 25 mJ for neon), the signal (1350 nm, 3 mJ), and the idler (2050 nm, 1.5 mJ) \cite{xue_fully_2020}. The combined optical pulse of the pump, signal, and idler (waveform-synthesized pulse) are focused by two lenses (focal length: 4 m for the pump and 3.5 m for the signal and idler) and propagating co-linearly into the gas cell (8- and 4-cm long for Ar and Ne, respectively). A weak gating pulse (800 nm, 30 fs) for AO-FROG was sampled using a similar configuration in Ref. \cite{xue_gigawatt-class_2022} but without polarization conversion, and it propagated noncollinearly with a small cross angle (~3 mrad). The focal point of all beams is set in front of the entrance of the gas cell to maintain short-trajectory phase matching \cite{salieres_coherence_1995,balcou_generalized_1997}. The relative delay time between the synthesizer channel and the gating channel was stabilized and controlled by a homemade stabilization circuit based on a He–Ne laser-injected Mach–Zehnder interferometer, which suppressed the delay jitter to less than 56 as \cite{xue_gigawatt-class_2022}.

To efficiently generate a high-harmonic beam, we optimized the waveform synthesizer parameters. The driving intensity was finely tuned to maximize the HH intensity by shifting the focal position before the inlet of the gas cell. The target gas was statically filled in the cell with Ar or Ne gas with the optimized gas pressure of approximately 3 or 10 torr to satisfy the phase matching condition, respectively. The synthesized field intensity for the HHG in Ar along the interaction area was estimated to $1\times10^{14} \rm{W/cm^{2}}$, with a peak power intensity ratio of 0.9:0.1:0.02 for the pump, signal, and idler, respectively. To obtain IAPs, a zirconium (Zr) or argentum (Ag) metal filter, which extracts the cutoff and continuum region, was inserted into the optical path at 4.5 m behind the gas cell. The extracted HH was then spectrally resolved by aberration-corrected concave grating, illuminated on a fluorescent screen after a microchannel plate, and recorded by a CCD camera.

\begin{figure}[b!]
\centering
\includegraphics[width=\linewidth, keepaspectratio]{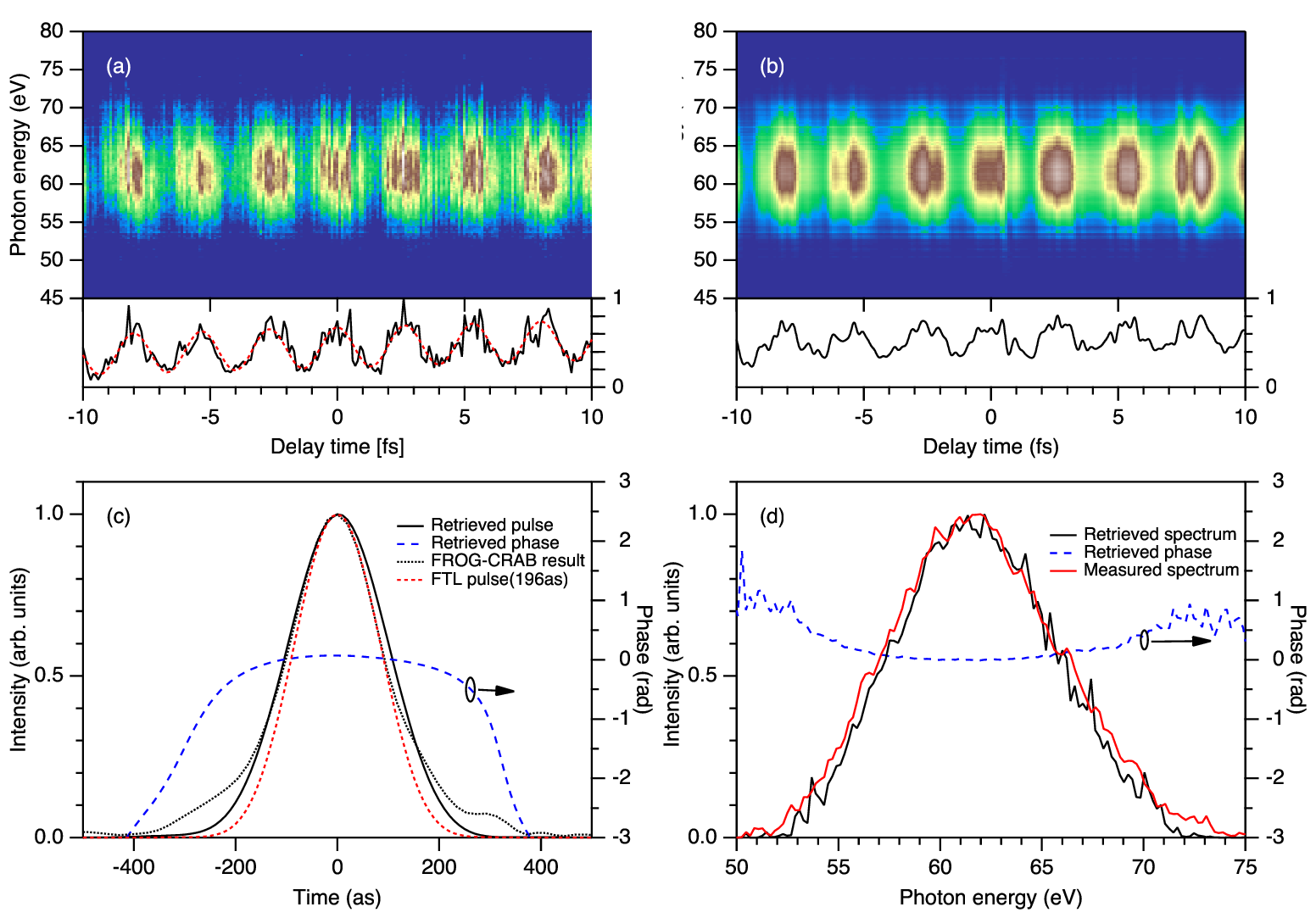}
\caption{Results of gigawatts-scale extreme ultraviolet isolated attosecond pulses (IAPs) in Ar gas. (a) Measured AO-FROG trace. The lower curve shows the modulation of the intensity at 60 eV, and the fitting result (red). (B) Reconstructed trace with an RMS error of 1.89\%. The lower black line shows the retrieved gating field. (c) Reconstructed IAP, black line: temporal intensity with a full width at half maximum (FWHM) duration of 227 as; blue dash line: temporal phase; red dash line: FTL pulse; black dot line: FROG-CRAB result. (d) Reconstructed IAP spectrum, black line: spectral intensity; blue dash line: spectral phase; red line: measured spectrum.}\label{fig2}
\end{figure}

Fig. \ref{fig1}(a) shows the calibrated single-shot spectrum (red dashed line) with an extended cutoff region of approximately 70 eV. We optimized the relative phase in the waveform synthesizer to produce a significant continuum structure at approximately 70 eV. To maintain stable supercontinuum generation (Fig. \ref{fig1}(a), red line: 100 shots spectrum filtered by 100-nm Zr), both the CEP and relative delay time were locked with the same device used in previous studies \cite{xue_fully_2020,xue_gigawatt-class_2022}. Fig. \ref{fig1}(b) shows the stability performances. In contrast to the free-run situation, the phase jitter of the pump and the delay jitter between the three channels were well stabilized in the system, which is consistent with a previous report \cite{xue_custom-tailored_2021}. By employing a FROG-like gating mechanism, the set of the HH spectra as a function of the relative delay between the synthesized pulse and the gate pulse indicates an electron recombination phase modulated by the gating field. As shown in Fig. \ref{fig2}(a), the measured trace consists of 201 spectral columns corresponding to different relative delays of the gating pulse with 100 attosecond step filtered by a 100-nm-thick Zr foil. The measured trace exhibited delay-dependent modulation with an oscillating cycle of approximately 2.7 fs, which corresponds to the electric field cycle of the gating pulse (800 nm, 1 cycle = 2.67 fs). Although the laser repetition was only 10 Hz, because of the intense HHG and simple optical setup, an exposure of 1 s for each delay step was sufficient to obtain a high SNR (even a single shot in theory). The proposed method significantly reduced the FROG trace acquisition time to only a few minutes compared with that of the conventional 10-Hz FROG-CRAB method \cite{xue_gigawatt-class_2022}. A homemade iteration program based on the least-squares generalized projection  algorithm \cite{gagnon_accurate_2008} was developed to retrieve temporal and spectral information. Figure \ref{fig2}(b) shows the reconstructed trace with a minimized root-mean-square error of 1.89\% after several iterations. As shown in Fig. \ref{fig2}(c), an IAP with a full width at half maximum (FWHM) of ~227 as centered at 60 eV was retrieved. Moreover, because only the cutoff region over 50 eV was extracted by the metal filter, the relatively low dispersion/atto-chirp indicated by the retrieved spectral phase (Fig. \ref{fig2}(d)) showed less atto-chirp in the cutoff region \cite{mairesse_optimization_2004,varju__frequency_2005,lopez-martens_amplitude_2005}. To confirm the accuracy of the proposed AO-FROG, we compared the same result with that using FROG-CRAB. The temporal profile (black dot line) of the reconstructed pulse (Fig. \ref{fig2}(c)), has almost the same duration as that obtained using AO-FROG, indicating that the proposed AO-FROG is reliable.

\begin{figure}[b!]
\centering
\includegraphics[width=\linewidth, keepaspectratio]{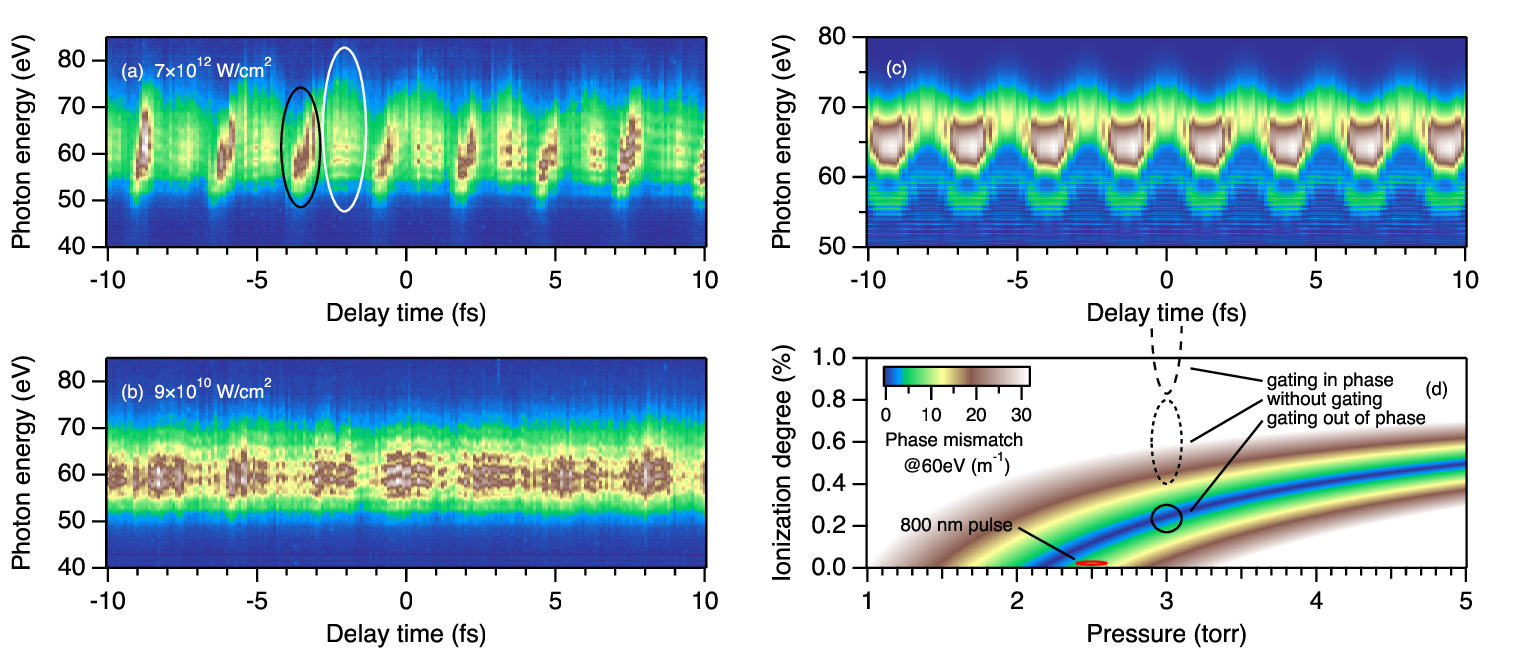}
\caption{Gating-field intensity dependence of AO-FROG traces, with a gating-field intensity of (a) $7\times10^{12} \rm{W/cm^{2}}$ and (b) $9\times10^{10} \rm{W/cm^{2}}$. (c) simulated trace with $2\times10^{12} \rm{W/cm^{2}}$ intensity. (d) phase matching condition of different synthesized fields.}\label{fig3}
\end{figure}

Notably, an adequate intensity of the gating field is required to satisfy the perturbative approximation expressed in Equations (1) and (2), which is restricted by two conditions \cite{dudovich_measuring_2006,smirnova_coulomb_2006}: (a) $A_p\Delta x\ll1$ and (b) $A_p/(\omega\Delta x)\ll1$, where $\Delta x\sim(2I_p)^{-1/2}$ is the spatial extension of the ground state. The influence of the gating field near the core as electrons move is negligible under these two conditions. The proper gating-field peak power intensity is limited to less than $\sim 3\times10^{11} \rm{W/cm^{2}}$ to satisfy the perturbative conditions for the experimental parameters.

\begin{figure}[t!]
\centering
\includegraphics[width=\linewidth, keepaspectratio]{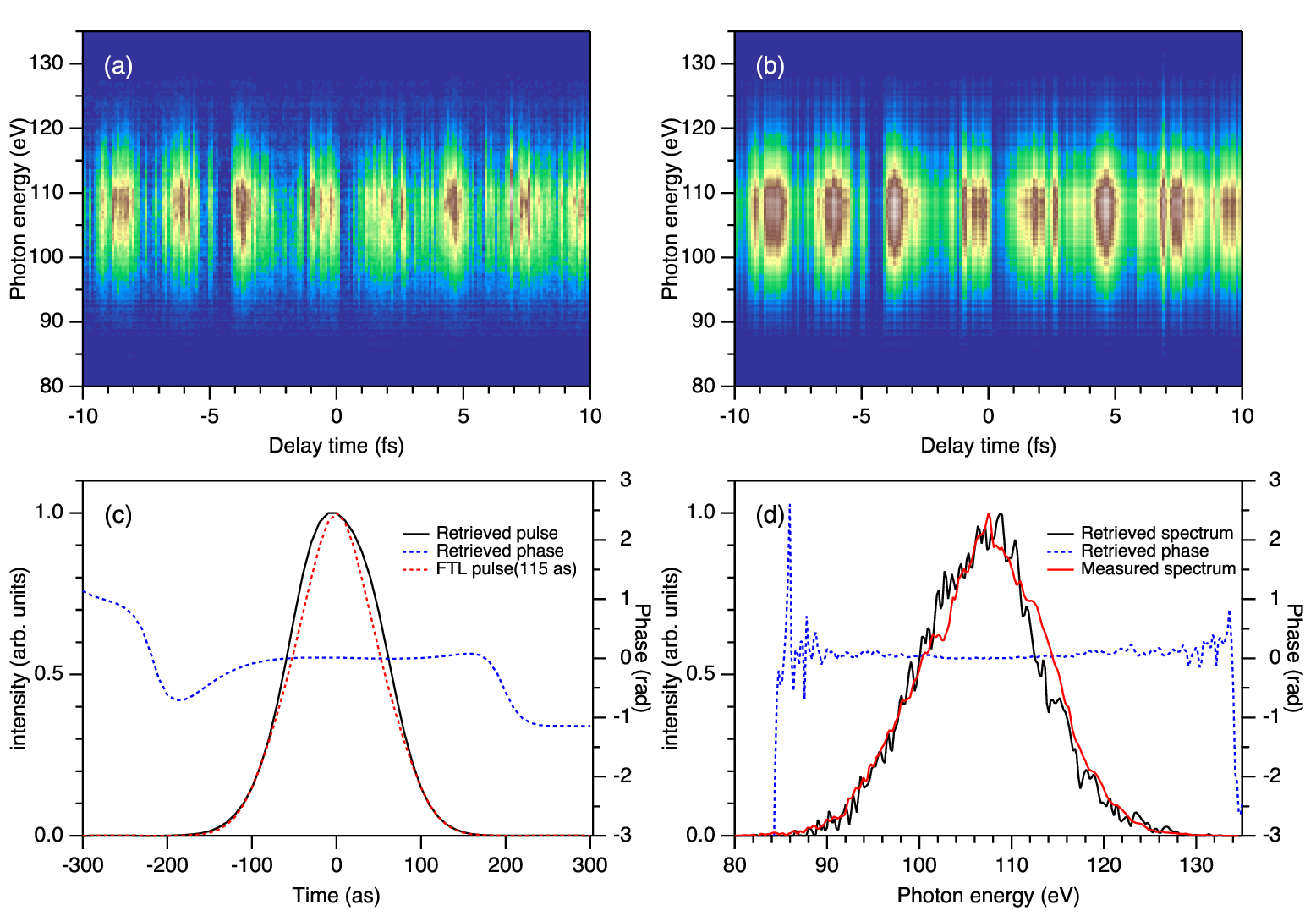}
\caption{Experimental result for IAPs generated in Ne gas centered at 107 eV. (a) Measured AO-FROG trace; (b) AO-FROG trace reconstructed using the least-squares generalized projection (LSGP) algorithm, with a reconstruction error of 2.52\%; (c) Reconstructed IAP profile: black line, temporal intensity with a FWHM duration of 128 as; blue dash line, temporal phase; red dash line: FTL pulse. (d) Reconstructed IAP spectrum profile: black line, spectral intensity; blue dash line, spectral phase; red line: measured spectrum.}\hrule\label{fig4}
\end{figure}

To confirm the influence of the gating-field intensity, we performed AO-FROG at different gating-field intensities. The measured AO-FROG traces are shown in Figs. \ref{fig3}(a) and \ref{fig3}(b). When the gating pulses with different intensities were coupled with the driving pulse and harmonics were generated, the obtained traces exhibited modulation structures different from those obtained at a gating-field peak intensity of approximately $2\times10^{11} \rm{W/cm^{2}}$ (Fig. \ref{fig2}(a)) with a clear periodic modulation. Fig. \ref{fig3}(a) shows a dense modulation with two patterns alternating with an interval of approximately 1.3 fs, which corresponds to the half cycle of the gating field. This modulation indicates that the strong gating field $(>10^{12} \rm{W/cm^{2}})$ surpasses the perturbative approximation and acts as a fourth channel in the waveform synthesizer. We infer that the strong gating field modified the synthesized electric field, thereby changing the ionization degree. The Ar gas pressure is firstly set to around 2.5 torr for better phase matching for pump driving field only (Fig. \ref{fig3}(d) red circle hint). With the help of signal and idler channel, the ionization degree increasing and the phase matching pressure was increased to around 3 torr for achieving brightest cut-off region (Fig. \ref{fig3}(d) black dot circle hint). The pressure is lower than the perfect phase matching pressure because it is necessary to balance the phase matching and reabsorption. When the gating field is in phase with the synthesized pulse, the ionization degree and cutoff photon energy increase because of the enhanced total electrical field. As shown in Fig. \ref{fig3}(a), the spectrum in the white-circled region is wider, but its intensity is weaker (phase matching deterioration, Fig. \ref{fig3}(d) black dash circle hint). The spectrum is also discontinuous because the gating field strengthens the side peaks of the synthesized field, resulting in attosecond pulse trains. However, when the two fields were out of phase, the electrical field was suppressed, and the ionization rate approached the optimal level (well phase matching under a fixed gas pressure, Fig. \ref{fig3}(d) black solid circle hint). This explains the narrower but more intense continuum in Fig. \ref{fig3}(a) (black circled region). The interplay between these two modes produced the modulation pattern shown in Fig. \ref{fig3}(a), corresponding to the alternating peaks with a period of approximately 1.3 fs. Fig. \ref{fig3}(c) shows the simulated AO-FROG trace using SAF theory, which confirms the explanation for the half-cycle modulation. In contrast, Fig. 3(b) shows that under a weak gating field $(<10^{11} \rm{W/cm^{2}})$, the modulation nearly disappeared. In this case, the effect of the gating field on the synthesized driving field and the ionization degree can be neglected, and the spectral modulation induced by the gating field becomes unclear compared with that in Fig. \ref{fig2}(a). These results demonstrate the reason for the half-cycle modulation under a strong gating field and guide the adjustment of the gating field intensity during experiments.

The experiments in Ar gas confirmed that AO-FROG can produce correct attosecond pulse durations. Compared with conventional FROG-CRAB, the all-optical approach does not require a central momentum approximation assumption or an e-TOF-based complicated setup, making it more spectral-flexible. Especially for attosecond pulse characterization at higher photon energies, it is difficult to perform FROG-CRAB measurements due to the lack of efficient XUV optics. Hence, we performed AO-FROG in the soft X-ray region where the photon energy exceeded 100 eV. To achieve a high-photon-energy attosecond pulse, Ne gas was employed as the interaction medium. The waist position of the synthesized waveform pulse was moved close to the inlet of the gas target to reach a smaller focus spot with an intensity of approximately $2.5\times10^{14} \rm{W/cm^{2}}$. The interaction medium was shortened to 4 cm to suppress the absorption of the soft X-ray radiation \cite{constant_optimizing_1999}. The Ne gas pressure was optimized to 10.6 torr for better phase matching. The blue dashed line in Fig. \ref{fig1}(b) indicates the 10-shot spectrum for Ne. The cutoff region of HH was extended to approximately 120 eV. Subsequent experimental procedures were similar to those in the case of Ar, but the metal filter was replaced with a 100-nm-thick Ag film for a high transmission rate above 100 eV. The blue line in Fig. \ref{fig1}(b) shows the extracted 100-shot spectrum. The same gating-field intensity ($\sim2\times10^{11} \rm{W/cm^{2}}$) was applied, and the AO-FROG trace was recorded (Fig. \ref{fig4}(a)). Continuum harmonics with high photon energy were generated, and the HH signal was less than that obtained when Ar was used, which is attributed to the low conversion efficiency and small transmission rate (<3\%) of the Ag film around 100 eV. An average of only 20 shots was required to achieve sufficient SNR for the retrieval process per step, corresponding to approximately 7 min of data acquisition for each scan. The reconstructed trace shown in Fig. \ref{fig4}(b) has a mean square error of 2.52\%. Figs. \ref{fig4}(c) and (d) show the retrieved temporal and spectral information, respectively. We obtained a supercontinuum of 90 to 120 eV, which corresponds to an IAP duration of 128 as in the FWHM. According to the estimated several tens-of-nanojoule energy of continuum harmonics \cite{xue_custom-tailored_2021}, these results demonstrate the generation of a subgigawatt-scale (>0.1 GW) IAP with a center photon energy of 107 eV. The obtained peak power is higher than that reported for soft X-ray IAPs. These results also demonstrate the spectral flexibility of AO-FROG.

In conclusion, we developed a perturbed waveform synthesizer for generating and characterizing intense, low-repetition-rate IAPs. The proposed approach was effective in analyzing IAPs, achieving results that closely match those of FROG-CRAB while offering improved spectral flexibility and shorter feedback cycles, as revealed by the temporal characterization of subgigawatt IAP (128 as, 107 eV, several tens of nanojoules). Although the repetition rate of the system was only 10 Hz, the data acquisition time for the 20-shot accumulation of the AO-FROG traces was only a few minutes. In principle, the data acquisition time can be further reduced by employing single-shot measurements. For a tunable IAP source, we developed a simplified, rapid, and robust approach that provides real-time feedback to the waveform synthesizer system, significantly enhancing the precise tuning of the attosecond pulse durations, thereby promoting the future application of intense attosecond pulses. In addition, The proposed method is a promising alternative to FROG-CRAB in cases where efficient soft X-ray optics are not available, e.g., octave-spanning supercontinuum \cite{nishimiya_octave-spanning_2024} driven by a single-cycle laser \cite{xu_dual-chirped_2024}. The proposed gigawatt/subgigawatt attosecond sources will pave the way for the intense attosecond science frontier, thereby promoting the research and development of ultrafast phenomena and nonlinear optics.

\begin{backmatter}
\bmsection{Funding}This work was funded by the Ministry of Education, Culture, Sports, Science and Technology of Japan (MEXT) through the Quantum Leap Flagship Program (Q-LEAP) (grant no. JP-MXS0118068681) and the RIKEN TRIP initiative (Leading-edge semiconductor technology).
\bmsection{Acknowledgments}D. D. thanks the International Program Associate program of RIKEN. Y. F. and H. W. acknowledge the National Key Research and Development Program of China (2022YFE0111500) and the International Partnership Program of CAS (115GJHZ2023023FN).B. X. acknowledges the CAS Pioneer Hundred Talents Program).
\end{backmatter}







\end{document}